\documentclass[aps,prl,twocolumn,reprint,amsmath,amssymb,superscriptaddress]{revtex4-2}

\usepackage{bm}
\usepackage{booktabs}
\usepackage[hidelinks]{hyperref}

\newcommand{\sltwo}{\mathrm{SL}(2,\mathbb{R})}
\newcommand{\slz}{\mathrm{SL}(2,\mathbb{Z})}
\newcommand{\sotwo}{\mathrm{SO}(2)}
\newcommand{\slalg}{\mathfrak{sl}(2,\mathbb{R})}
\newcommand{\so}{\mathfrak{so}(2)}
\newcommand{\tr}{\mathrm{tr}}
\newcommand{\Mten}{M_{10}}
\newcommand{\Mel}{M_{11}}

\newcommand{\cV}{{\cal V}}
\newcommand{\dd}{{\rm d}}

\begin{document}

\title{Type IIB Forms and Scalars from an $\sltwo$-invariant 
\\[3pt]
Topological Field Theory
with Boundary}

\author{Euihun Joung}
\email{euihun.joung@khu.ac.kr}
\affiliation{Department of Physics and Research Institute of Basic Science, Kyung Hee University, Seoul 02447, Republic of Korea}

\author{J.~A.~Rosabal}
\email{jarosabal80@khu.ac.kr}
\affiliation{Department of Physics and Research Institute of Basic Science, Kyung Hee University, Seoul 02447, Republic of Korea}

\begin{abstract}
We formulate the scalar and form-field sector of type IIB supergravity as an eleven-dimensional topological field theory with boundary. Unlike the pseudo-action, it is a genuine action principle with manifest global $\sltwo$ symmetry. The IIB fields arise as edge modes: a boundary term breaks the $\sltwo$ gauge symmetry to $\sotwo$, so the axion--dilaton parametrizes $\sltwo/\sotwo$, and the boundary field equations give the duality relations, including $F_5$ self-duality. As the action never refers to the axion--dilaton directly, S-duality monodromies become holonomies of a flat connection.
\end{abstract}

\maketitle

\subsection{Introduction} 

Not every physical system is best described by a standard Lagrangian in its own dimension. Chiral scalars~\cite{Floreanini:1987as} and chiral, duality-invariant form fields~\cite{Henneaux:1988gg,Schwarz:1993vs} are naturally realized on the boundary of a topological field theory in one higher dimension~\cite{Witten:1996hc,Witten:1999vg,Hopkins:2002rd,Belov:2006jd,Freed:2012bs,Arvanitakis:2022bnr,Chen:2025higher,Heckman:2017fqh}. The bulk has no local degrees of freedom but provides a manifestly covariant action principle in which duality relations become boundary equations of motion. 

This approach has yielded democratic actions for the form fields of eleven-dimensional supergravity~\cite{Evnin:2023ypu,Rosabal:2025} (see also Ref.~\cite{Banerjee:2026sym} for an application to symmetry descent and cohomotopy), for the RR fields of type II supergravities~\cite{Evnin:2023ypu}, whose duality relations must be imposed by hand in the standard pseudo-actions~\cite{Bergshoeff:2001pv}, and for linearized gravity~\cite{Chen:2026lin}.

Electric--magnetic duality is, however, only part of the duality structure of supergravity: the scalars parametrize a coset $G/H$, on which a duality group $G$ acts nonlinearly while rotating the form fields linearly, as in S- and U-duality~\cite{Hull:1994ys}. Scalars on a coset $G/H$ are also familiar as edge modes of topological gauge theories, from coset models on the boundary of Chern--Simons theory~\cite{Moore:1989yh} to BF theories with gauge group $G$ whose boundary breaks $G$ to $H$~\cite{Bonetti:2024symtft,Antinucci:2024ssb,Apruzzi:2025spacetime,Bonetti:2025nonlinear}. To our knowledge, this description has not yet been combined with the electric--magnetic duality of the form fields, as supergravity requires. For type IIB, existing topological constructions treat the $p$-form sector~\cite{Apruzzi:2023uma} or the RR sector on a fixed NSNS background~\cite{Belov:2006xj}; in these, the scalars enter at most through the axion pair $(F_1,F_9)$, and the dilaton at most as a background. In the spirit of F-theory, type IIB has also been conjectured to be an edge mode of a twelve-dimensional topological bulk~\cite{Heckman:2017fqh}. Ten-dimensional formulations with manifest $\sltwo$ exist as actions with auxiliary fields~\cite{DallAgata:1998va,Mkrtchyan:2022ulc} or as democratic pseudo-actions~\cite{Fernandez-Melgarejo:2023upq}.

Type IIB is the prototypical test case: its self-dual five-form has a vanishing kinetic term, so the theory has traditionally been defined through its field equations~\cite{Schwarz:1983qr,Howe:1983sra}; covariant actions require auxiliary fields~\cite{DallAgata:1997ju,DallAgata:1998va,Sen:2015nph,Avetisyan:2022zza,Mkrtchyan:2022ulc,Hull:2023dgp,Sen:2019qit,Chakrabarti:2025gyt,Chakrabarti:2026uwa}; and the pseudo-action, in which self-duality is imposed only after variation, neither defines a path integral by itself nor reproduces the holographic on-shell action~\cite{Kurlyand:2022vzv, Adhikari:2026iib}.
S-duality acts as an $\sltwo$ symmetry on the axion--dilaton $\tau$ and the two-forms. Existing formulations describe the scalars through $\tau$ or the coset matrix $M\in\sltwo/\sotwo$, only local sections of a bundle with transition functions in the duality group. In S-folds~\cite{Garcia-Etxebarria:2015wns,Aharony:2016kai}, J-folds~\cite{Inverso:2016eet,Assel:2018vtq}, and F-theory backgrounds~\cite{Vafa:1996xn}, $\tau$ returns to itself only up to $\slz$, so no global section exists and the Lagrangian is defined only for a monodromy fixed in advance.

We show that combining the two topological descriptions resolves both problems. The key point, valid for any coset $G/H$, is that the scalars are described not by a coset representative but by a flat $G$ connection $A_1$~\cite{Bonetti:2024symtft,Bonetti:2025nonlinear}, globally defined for any monodromy, to which the form fields couple in representations of $G$. For type IIB, $G=\sltwo$, and our main result is the action 
\begin{equation}
    S_{\rm bTFT}=S_{\rm TFT}+S_{\rm b}
\end{equation}
of a topological field theory with boundary (bTFT) on an eleven-manifold $\Mel$ with boundary $\partial\Mel=\Mten$, whose bulk and boundary parts are
\begin{align}
S_{\rm TFT}=\frac i2\int_{\Mel}\Big\{
&\tr\big[A_1\,\dd A_9+A_9\,\dd A_1+2\,A_1{}^2A_9\big]
\nonumber\\
&+A_3\,\dd A_7+\dd A_3\,A_7+2\,A_3A_1A_7
\nonumber\\
&+A_5\,\dd A_5+A_3\,\epsilon\,A_3^tA_5\Big\}\,,
\label{eq:S11}\\[4pt]
S_{\rm b}=\frac{1}{4}\int_{\Mten}\Big\{
&\tr\big[2\,A^{\mathbb S}_1*A^{\mathbb S}_1+\tfrac{1}{2}\,A^{\mathbb S}_9*A^{\mathbb S}_9\big]
\nonumber\\
&+A_3*A_3^t+A_7^t*A_7+A_5*A_5\Big\}\,,
\label{eq:S10}
\end{align}
where wedge products are implicit and $X^{\mathbb S}=\frac{1}{2}(X+X^t)$. Here $(A_1)_\alpha{}^\beta$ and $(A_9)_\alpha{}^\beta$ are $\slalg$-valued, $(A_3)^\alpha$ and $(A_7)_\alpha$ are row and column doublets, $A_5$ is a singlet, and $\epsilon$ is the Levi-Civita symbol. On the boundary, the forms $A_p$ reduce to the type IIB field strengths.

\subsection{Type IIB supergravity}

In $\sltwo$-covariant form, the bosonic pseudo-action of the scalar and gauge-field sector~\cite{Note1} reads
\begin{align}
S_{\rm pseudo}=\int_{\Mten}\Big[&\tfrac{1}{4}\,\Omega^i{}_j\,*\Omega^j{}_i+\tfrac{1}{2}\, M_{ij}\,H_3^i\,*H_3^j\nonumber\\
&+\tfrac{1}{4}\,F_5\,*F_5+\tfrac{i}{4}\,\epsilon_{ij}\,D_4\,H_3^i\,H_3^j\Big],
\label{eq:SE}
\end{align}
with $\Omega^i{}_j=M^{ik}\dd M_{kj}$, $H_3^i=\dd B_2^i$ and $F_5=\dd D_4-\frac{1}{2}\,\epsilon_{ij}\,B_2^i\,\dd B_2^j$. The symmetric, unit-determinant matrix $M=e^{\phi}\left(\begin{smallmatrix}C_0^2+e^{-2\phi}&-C_0\\-C_0&1\end{smallmatrix}\right)$ encodes the axion $C_0$ and the dilaton $\phi$, and $B_2^i=(B_2,C_2)$ is the NSNS--RR doublet. Introducing dual field strengths through the duality relations
\begin{equation}
*\Omega^i{}_j=-i\,F_9{}_j{}^i\,,\quad\ M_{ij}\,*H_3^j=-i\,H_{7i}\,,\quad\ *F_5=-i\,F_5\,,
\label{eq:dual}
\end{equation}
turns the field equations into Bianchi identities,
\begin{align}
\dd\Omega^i{}_j+\Omega^i{}_k\,\Omega^k{}_j&=0,\nonumber\\
\dd H_3^i&=0,\nonumber\\
\dd F_5+\tfrac{1}{2}\,\epsilon_{ij}\,H_3^i\,H_3^j&=0,\nonumber\\
\dd H_{7i}+\epsilon_{ij}\,H_3^j\,F_5&=0,\nonumber\\
\dd F_9{}_j{}^i-H_3^i\,H_{7j}+\tfrac{1}{2}\,\delta^i_j\,H_3^k\,H_{7k}&=0.
\label{eq:bianchi}
\end{align}
The first line holds identically for $\Omega=M^{-1}\dd M$, which is also traceless since $\det M=1$.
By \eqref{eq:dual}, the last line is the dual of the field equation for $M$. This equation has only two independent components; correspondingly, $F_{9j}{}^i$ satisfies the linear constraint $\epsilon^{ij}M_{jk}(F_9)_i{}^k=0$~\cite{DallAgata:1998va,Bergshoeff:2005ac}.

\subsection{Topological bulk}

The first ingredient of the action \eqref{eq:S11}--\eqref{eq:S10} turns electric--magnetic duality into a symplectic structure. Democratic type IIB supergravity has exactly one $\sltwo$ multiplet of field strengths in each odd degree from one to nine, the $\sltwo$ content of its tensor hierarchy~\cite{Bergshoeff:2005ac}, and in eleven dimensions a $p$-form and a $(10-p)$-form are canonically conjugate through $\int A_p\,\dd A_{10-p}$. The bulk therefore contains one form for each type IIB field strength, of the same degree and $\sltwo$ representation. 
Schematically, the duality pairs $(\Omega,F_9)$ and $(H_3,H_7)$ correspond to the symplectic pairs $(A_1,A_9)$ and $(A_3,A_7)$, while the self-dual five-form is paired with itself through $A_5$.
Hence the kinetic terms of $S_{\rm TFT}$ are democratic, and it needs no metric.

For the scalars, writing $M=\cV\,\cV^t$, S-duality acts on the representative $\cV$ from the left, $\cV\to\gamma\cV$, while $\cV$ is defined only up to local $\sotwo$ rotations from the right. In the bulk, this local $\sotwo$ is enlarged to a local $\sltwo$, acting as $\cV\to\cV g$, with $A_1$ as its connection. 

The bulk equations of motion of $S_{\rm TFT}$, which we refer to as Gauss laws, are
\begin{align}
\dd A_1+A_1\,A_1&=0\,,\nonumber\\
\dd A_3+A_3\,A_1&=0\,,\nonumber\\
\dd A_5+\tfrac{1}{2}\,A_3\,\epsilon\,A_3^t&=0\,,\nonumber\\
\dd A_7+A_1\,A_7+\epsilon\, A_3^t\,A_5&=0\,,\nonumber\\
\dd A_9+A_1\,A_9+A_9\,A_1+(A_7\,A_3)_0&=0\,,
\label{eq:bulk}
\end{align}
where $(X)_0=X-\frac{1}{2}(\tr X)\,\mathbf 1$ denotes the traceless part. Each arises from varying the symplectic partner of the field it constrains; for instance, the flatness of $A_1$, the equation of motion of $A_9$, implies the first line of \eqref{eq:bianchi}, an identity in supergravity.
The system is integrable: $\dd$ of each Gauss law vanishes on the solution space.

The bulk action is invariant, up to boundary terms, under local $\sltwo$ transformations,
\begin{align}
A_1&\to g^{-1}(\dd+A_1)\,g\,,\qquad A_3\to A_3\,g\,,\nonumber\\
A_7&\to g^{-1}A_7\,,\qquad A_9\to g^{-1}A_9\,g\,,
\label{eq:SL2gauge}
\end{align}
where $A_5$ is invariant and $g\,\epsilon\,g^t=\epsilon$, as well as under the higher-form gauge transformations of $A_3,\dots,A_9$.
Since each pair enters symmetrically, $S_{\rm TFT}$ differs from a gauge-invariant BF-type action by $-\tfrac i2\int_{\Mten}[\tr(A_1A_9)+A_3A_7]$, so that under \eqref{eq:SL2gauge}, $S_{\rm TFT}\to S_{\rm TFT}+\tfrac i2\int_{\Mten}\tr(A_9\,\dd g\,g^{-1})$.
Gauge transformations that do not vanish on $\Mten$ are therefore no longer redundancies: they act on the edge modes, the Goldstone modes of the gauge symmetry broken at the boundary.

\subsection{Boundary action}

These edge modes are precisely the type IIB fields, and a boundary action supplies their dynamics. For the five-form, this boundary term is essentially unique.
For the doublets, it requires a symmetric tensor, which $\sltwo$ does not preserve, so the form-field kinetic terms necessarily break $\sltwo$. Any positive-definite choice $s_{\alpha\beta}$ can be brought to $\delta_{\alpha\beta}$ by $\sltwo$, up to a scale fixed by matching the IIB fields; 
its stabilizer $\sotwo$ is thus fixed by the form fields themselves, whereas in the coset constructions of Refs.~\cite{Apruzzi:2025spacetime,Bonetti:2025nonlinear} the breaking pattern is part of the boundary data.
 We require the remaining boundary terms and conditions to preserve this $\sotwo$ locally, as in the coset formulation. Under $h\in\sotwo$, conjugation preserves the split $X=X^{\mathbb S}+X^{\mathbb A}$ into symmetric and antisymmetric parts and $h^{-1}\dd h$ is antisymmetric, so $A_1^{\mathbb S}\to h^{-1}A_1^{\mathbb S}h$ is covariant, while $A_1^{\mathbb A}=a\,\epsilon$ transforms as an abelian gauge field, $a\to a+\dd\theta$ for $h=e^{\theta\epsilon}$. A boundary term quadratic in $a$ is not invariant under this shift, so only $A_1^{\mathbb S}$ may appear. 
 On the bulk side, the variation reduces to $\tfrac i2\int_{\Mten}\tr(A_9^{\mathbb A}\,\dd h\,h^{-1})$, so the Dirichlet condition $A_9^{\mathbb A}|_{\Mten}=0$, analogous to the coset boundary conditions of Refs.~\cite{Apruzzi:2025spacetime,Bonetti:2025nonlinear}, ensures local $\sotwo$ invariance. Neither $A_1^{\mathbb A}$ nor $A_9^{\mathbb A}$ then appears in $S_{\rm b}$, and $A_1^{\mathbb A}$ is constrained by $\dd A_1^{\mathbb A}+A_1^{\mathbb S}A_1^{\mathbb S}=0$.
 The metric enters only through $S_{\rm b}$, and on the boundary equations of motion the energy--momentum tensor is that of the IIB fields.

Varying the full action $S_{\rm bTFT}$ with all boundary values free except $A_9^{\mathbb A}|_{\Mten}=0$ gives one boundary equation for each symplectic pair,
\begin{equation}
*A_1^{\mathbb S}=-\tfrac{i}{2}\,A_9^{\mathbb S}\,,
\qquad *A_3^t=-i\,A_7\,,\qquad *A_5=-i\,A_5
\,.
\label{eq:bcproj}
\end{equation}
For $(A_1,A_9)$ and $(A_3,A_7)$, the coefficients in $S_{\rm b}$ ensure that varying either member gives the same equation, up to Hodge duality. $S_{\rm b}$ vanishes when they hold~\cite{Note2}.
As in supergravity, only the coset part of the scalar current is dualized. Its $\so$ part, $A_1^{\mathbb A}$, acts as the composite $\sotwo$ connection, so exactly two scalars remain. The Dirichlet condition is the bulk counterpart of the linear constraint below \eqref{eq:bianchi} and is compatible with the bulk equations: once \eqref{eq:bcproj} holds, the antisymmetric part of the last equation in \eqref{eq:bulk} is satisfied identically.

\subsection{Boundary reduction}

Apart from the five-form, $S_{\rm bTFT}$ is a BF theory, with $A_7$ and $A_9$ as bulk Lagrange multipliers. Integrating them out imposes the first two Gauss laws in \eqref{eq:bulk}, while their boundary values are fixed by the first two equations in \eqref{eq:bcproj}. Locally, these conditions are solved by
\begin{equation}
A_1=\cV^{-1}\dd\cV\,,\qquad 
A_3=H_3\,\cV\,,
\label{eq:puregauge}
\end{equation}
with $\cV$ an $\sltwo$-valued function  and $H_3$ a row doublet of three-forms with $\dd H_3=0$.
Substituting \eqref{eq:puregauge}, the action becomes
\begin{align}
S_{\rm bTFT}=&\int_{\Mten}\Big[\tr\big((\cV^{-1}\dd\cV)^{\mathbb S}*(\cV^{-1}\dd\cV)^{\mathbb S}\big)\nonumber\\
&\qquad+\tfrac{1}{2}\,H_3\,\cV\cV^t*H_3^t+\tfrac{1}{4}\,A_5*A_5\Big]\nonumber\\
&+\tfrac{i}{2}\int_{\Mel}A_5\big(\dd A_5+H_3\,\epsilon\,H_3^t\big),
\label{eq:Sred}
\end{align}
where $A_5$ is kept as an independent field. The scalar and two-form terms come from the boundary term of $S_{\rm TFT}$, since those of $S_{\rm b}$ cancel on \eqref{eq:bcproj}.
Since $M^{-1}\dd M=2\,(\cV^t)^{-1}(\cV^{-1}\dd\cV)^{\mathbb S}\,\cV^t$, the first two terms reproduce the scalar and two-form terms of \eqref{eq:SE}. In components, $M_{ij}=\cV_i{}^\alpha\,\cV_j{}^\beta\,\delta_{\alpha\beta}$, so that the frame $\cV_i{}^\alpha$ converts the supergravity doublet indices $i,j$ into the frame indices $\alpha,\beta$.

The five-form can be reduced further, but only on shell. For $A_5=F_5$ its kinetic term matches that of \eqref{eq:SE}, and its Gauss law, $\dd F_5+\frac{1}{2}\,H_3\,\epsilon\,H_3^t=0$, agrees with the definition of $F_5$. The bulk term in \eqref{eq:Sred} then reduces to the Chern--Simons coupling $\frac{i}{4}\int_{\Mten}D_4\,H_3\,\epsilon\,H_3^t$, giving the pseudo-action \eqref{eq:SE}, which no longer implies self-duality. 

Imposing the remaining Gauss laws reduces the boundary fields further on shell: 
\begin{align}
A_1&=\cV^{-1}\dd\cV\,,\qquad A_3=H_3\,\cV\,,\qquad A_5=F_5\,,\nonumber\\
A_7&=\cV^{-1}H_7\,,\qquad A_9=\cV^{-1}F_9\,\cV\,.
\label{eq:dict}
\end{align}
They become the components of the type IIB field strengths in the frame $\cV$.
The last entry of \eqref{eq:dict} is symmetric because $A_9^{\mathbb A}$ vanishes on $\Mten$. On $\Mten$, the Gauss laws \eqref{eq:bulk} then reduce, line by line, to the Bianchi identities \eqref{eq:bianchi}, and \eqref{eq:bcproj} reduces to the duality relations \eqref{eq:dual}.

\subsection{S-duality monodromy as a Wilson loop}

The representative $\cV$ exists only locally and enters only through the solution \eqref{eq:puregauge}, not through $S_{\rm bTFT}$ or its equations of motion. Since S-duality acts on $\cV$ from the left, its action on the IIB fields in \eqref{eq:dict} is compensated by that on $\cV$, leaving every $A_p$ invariant; for example, $A_3=H_3\,\cV\to H_3\,\gamma^{-1}\gamma\,\cV=A_3$.
Every $A_p$ therefore remains globally defined on $\Mten$, even when $\cV$ and $B_2$ are glued across patches by S-duality, and the holonomy of $A_1$ is conjugate to the monodromy, the transition functions of $\cV$.
Classically, configurations related by a global $\sltwo$ transformation are therefore identified: S-duality becomes a redundancy, and the asymptotic value of $\tau$ is no longer a parameter. In string theory only $\slz$ is a redundancy, so this value remains a parameter modulo $\slz$, as we expect here once flux quantization or charged sources reduce $\sltwo$ to $\slz$.

J-folds~\cite{Inverso:2016eet,Assel:2018vtq} give an explicit example. They are supersymmetric type IIB solutions on $\mathrm{AdS}_4\times S^1_\eta\times M_5$, where $\eta\sim\eta+T$, and in eleven dimensions have $A_1=\tfrac12\,\mathrm{diag}(-a_1,a_1)$. The first Gauss law reduces to $\dd a_1=0$, and since $\oint_{S^1_\eta}a_1=-2\,T$, $a_1$ is closed but not exact. All other fields are independent of $\eta$, so the monodromy is carried entirely by the Wilson loop
\begin{equation}
\mathfrak M\equiv\mathrm P \exp\oint_{S^1_\eta} A_1=\begin{pmatrix}e^{T}&0\\0&e^{-T}\end{pmatrix}\,.
\end{equation}
To compare with supergravity, write $a_1=\dd\phi$ locally on $\Mten$. Then $\cV=\mathrm{diag}(e^{-\phi/2},e^{\phi/2})$, and $\phi=\phi_0-2\,\eta$ is the dilaton of the J-fold of Ref.~\cite{Inverso:2016eet} in the frame with vanishing axion. Since $\phi$ decreases by $2\,T$ around the circle, $\cV(\eta+T)=\mathfrak M\,\cV(\eta)$ and $B_2(\eta+T)=B_2(\eta)\,\mathfrak M^{-1}$, so $\tau$ returns to itself only up to $\mathfrak M$. In $A_3=\dd B_2\,\cV$ the two gluings cancel, and likewise for the other $A_p$. 
The monodromy therefore enters $S_{\rm bTFT}$ only through the holonomy of $A_1$, and nothing in the action is multivalued.

In string theory, the monodromy must also preserve the charge lattice: a constant frame change $P$ maps the lattice to $\mathbb Z^2$ and the monodromy to $P\,\mathfrak M\,P^{-1}=J_n=\left(\begin{smallmatrix}n&1\\-1&0\end{smallmatrix}\right)$, which requires $\tr\,\mathfrak M=2\cosh T=n\in\mathbb Z$ with $n\geq3$, as for the $J_n$-folds of Ref.~\cite{Assel:2018vtq}. This condition constrains $A_1$ alone, which is invariant under constant frame changes, while the lattice, encoded in $P$, fixes the $\slz$ class. If $S^1_\eta$ bounds a disc in $\Mel$, the Wilson loop must be trivial.

Beyond $\sltwo$, $S_{\rm TFT}$ is invariant under $\mathrm{SL}^\pm(2,\mathbb R)$ if $A_5\to(\det g)\,A_5$, and $S_{\rm b}$ under $\mathrm O(2)$. Reflections have no gauge field but can glue patches; around such a loop the doublets are reflected and $A_5$ flips sign, as under $(-1)^{F_L}$ or worldsheet parity~\cite{Debray:2021vob}, the monodromies of reflection seven-branes~\cite{Dierigl:2022reg}.

\subsection{Algebraic structure}

The fields assemble into a polyform $\Psi=A_1\oplus A_3\oplus A_5\oplus A_7\oplus A_9$ of total degree one, with $A_p$ valued in the degree-$(1-p)$ part of the Lie algebra $\mathfrak g=\slalg\oplus\mathbb R^2\oplus\mathbb R\oplus(\mathbb R^2)^*\oplus\slalg^*$, graded compatibly with its brackets. In terms of generators $X^\alpha{}_\beta$, $Q_\alpha$, $Z$, $P^\alpha$ and $Y^\alpha{}_\beta$ of the five components, the nonvanishing brackets, besides the $\slalg$ action on each component, are
\begin{equation}
[Q_\alpha,Q_\beta]=\epsilon_{\alpha\beta}\,Z\,,\quad [Z,Q_\alpha]=\epsilon_{\alpha\beta}\,P^\beta\,,\quad [P^\alpha,Q_\beta]=Y^\alpha{}_\beta\,.
\label{eq:brackets}
\end{equation}
The first encodes the cubic coupling. The second follows from it by invariance of the pairing $\langle\,,\rangle$, given by the trace and the natural contractions $\langle Q_\alpha,P^\beta\rangle=\delta_\alpha^\beta$ and $\langle Z,Z\rangle=1$. The last, the moment map of the doublets, sources the equation for $A_9$. Since all degrees are even, no signs arise, and $\mathfrak g$ is an ordinary Lie algebra, an extension of $\slalg$ by a nilpotent ideal.

The action $S_{\rm TFT}=\frac i2\int\langle\Psi,\dd\Psi\rangle+\frac i6\int\langle\Psi,[\Psi,\Psi]\rangle$ is then the higher Chern--Simons functional of $\mathfrak g$, and the Gauss laws \eqref{eq:bulk} are its Maurer--Cartan equation $\dd\Psi+\frac12[\Psi,\Psi]=0$. The Chevalley--Eilenberg algebra of $\mathfrak g$ is the free differential algebra of the type IIB field strengths~\cite{Castellani:1987,Castellani:1993ye} in its democratic form~\cite{Bergshoeff:2005ac}, which is why the Gauss laws reduce to the Bianchi identities \eqref{eq:bianchi}. In the $L_\infty$ language of Refs.~\cite{Sati:2008eg,Fiorenza:2013nha,Giotopoulos:2026cqts}, $\mathfrak g$ is a strict Lie $9$-algebra. What is new is that the scalars enter through the flat $\slalg$ current, so the pairing extends to the entire sector and turns this free differential algebra into the field equations of an action.

Modulo the dual components, the first bracket in \eqref{eq:brackets} makes $\mathbb R^2\oplus\mathbb R$ a Heisenberg algebra: the NSNS and RR gauge transformations of $A_3$ commute to a gauge transformation of $A_5$, which, when large, shifts the D3 Page charge~\cite{Maldacena:2001ss}, the algebraic origin of brane creation~\cite{Hanany:1996ie}.

The same structure encodes the Chern--Simons couplings in the doubled formalism of Ref.~\cite{Cremmer:1998px} (see also Ref.~\cite{Rosabal:2025}), where the subalgebra of $\mathfrak g$ with $\slalg$ restricted to its Borel part, which parametrizes the coset, already appeared. There, the equations of motion are a Maurer--Cartan equation supplemented by a twisted self-duality condition $*\mathcal G=\mathcal S\,\mathcal G$, where the twist $\mathcal S$, with $\mathcal S^2=-1$, relates each field strength to its dual.
Here, this condition is the boundary equation \eqref{eq:bcproj}, derived rather than imposed. 
Indeed, $S_{\rm b}=\frac i4\int_{\Mten}\langle\Psi,*\mathcal S\,\Pi\Psi\rangle$, where $\Pi$ projects out $\so\oplus\so^*$, and \eqref{eq:bcproj} with $A_9^{\mathbb A}|_{\Mten}=0$ is $\Pi(*-\mathcal S)\Psi|_{\Mten}=0$.
The twist is set by the symmetric metric $s_{\alpha\beta}$ in $S_{\rm b}$, here $\delta_{\alpha\beta}$; together with $\epsilon$, it defines a K\"ahler polarization of $(\mathbb R^2,\epsilon)$, which, dressed by $\cV$, becomes $M$. It also fixes $\Pi$: for general $s$, $X^{\mathbb S}$ becomes $\frac12(X+s\,X^t s^{-1})$.
Unlike that for linearized gravity~\cite{Chen:2026lin}, the twist here commutes only with $\so$. Such twisted boundary conditions, combined with an isotropic subalgebra, are known for Chern--Simons theory~\cite{Severa:2016lwc,Pulmann:2019vrw}.

Since $\mathfrak g$ is an ordinary Lie algebra, its gauge transformations $\delta\Psi=\dd\Lambda+[\Psi,\Lambda]$ close off shell. Finally, $S_{\rm TFT}$ extends to an AKSZ sigma model~\cite{Alexandrov:1995kv}, and the Lagrangian boundary relations \eqref{eq:bcproj} fit into the BV--BFV formalism~\cite{Cattaneo:2012qu}.

\subsection{Discussion}

The mechanism is not specific to type IIB. For any coset $G/H$, the local $H$ acting on $\cV$ is enlarged in the bulk to a local $G$ carried by the flat connection $A_1$, which the boundary term breaks back to $H$~\cite{Apruzzi:2025spacetime,Bonetti:2025nonlinear}, while the form fields, in representations of $G$, are paired by the bulk kinetic terms. In four dimensions, this pairing is the invariant symplectic form of the duality group~\cite{Gaillard:1981rj}, and for maximal supergravity~\cite{Cremmer:1979up} the construction should make electric--magnetic duality and its nonabelian extension $E_{7(7)}$ manifest together, with U-duality monodromies realized as holonomies.

The boundary terms in the variation of $S_{\rm TFT}$ define a symplectic form $\omega$ on the boundary values, pairing $A_1$ with $A_9$, $A_3$ with $A_7$, and $A_5$ with itself. Each relation in \eqref{eq:bcproj} defines a Lagrangian subspace $L$. For the duality relations, $L\cap\bar L=0$ and $\omega(v,\bar v)>0$ for $v\in L$: they are K\"ahler polarizations for the complex structure given by the Hodge star and the twist $\mathcal S$, real but still metric-dependent in Lorentzian signature. The condition $A_9^{\mathbb A}=0$ instead satisfies $L=\bar L$ and involves no metric: on the $\so$ sector, it is a real polarization of topological type. Other Lagrangian choices define other boundary theories for the same bulk, as in the SymTFT framework~\cite{Kapustin:2010hk,Freed:2012bs,Gaiotto:2020iye,Apruzzi:2021nmk,Heckman:2022xgu,Bonetti:2024symtft,Antinucci:2024ssb,Apruzzi:2025spacetime,Bonetti:2025nonlinear}. The boundary action $S_{\rm b}$, the only non-topological ingredient, thus does more than supply edge-mode dynamics: as the generating function of \eqref{eq:bcproj}, it selects the K\"ahler polarization.

$S_{\rm bTFT}$ also unifies the known ten-dimensional formulations as boundary reductions of a single theory. Integrating out fields requires a polarization: the paired fields admit a natural one, giving the conventional action, whereas the self-dual five-form, its own partner, admits no real covariant one and can be reduced only on shell, giving the pseudo-action. Solving the Hamiltonian constraints instead keeps every field off shell, including the scalars. A direct reduction should give non-covariant, duality-symmetric actions as in Refs.~\cite{Henneaux:1988gg,Schwarz:1993vs}, and a covariant one, along the lines of Ref.~\cite{Arvanitakis:2022bnr} (see also Ref.~\cite{Evnin:2023ypu}), actions with auxiliary fields as in Refs.~\cite{Mkrtchyan:2019opf,Bansal:2021bis,Avetisyan:2021heg,Avetisyan:2022zza,Evnin:2022kqn,Mkrtchyan:2022ulc}, from which the manifestly $\sltwo$-invariant PST-type~\cite{Pasti:1995ii,Pasti:1995tn,Pasti:1995us,Pasti:1996vs} action of Ref.~\cite{DallAgata:1998va} should follow.

In summary, $S_{\rm bTFT}$ is an action for type IIB, not a pseudo-action: the self-duality of $F_5$ and the S-duality monodromies are outputs of the variational principle. Compared with Refs.~\cite{Apruzzi:2023uma,Bonetti:2025nonlinear}, the new ingredient is the coupling of the nonabelian pair $(A_1,A_9)$ to the form fields, which cannot be truncated since the axion and dilaton components of $A_1$ do not separate. Brane sources should supply the charge lattice; with compact fields and this lattice, preserved by the holonomy of $A_1$, the quantum theory would reduce $\sltwo$ to $\slz$. 

The path integral on $\Mel$ defines a state on $\Mten$, which becomes a wave functional once a polarization is chosen; since the bulk is topological, the metric enters only through the K\"ahler polarization. As for the self-dual field~\cite{Witten:1996hc,Belov:2006xj}, this wave functional should reproduce the type IIB partition function as a section of a line bundle, now for the entire $\sltwo$-covariant sector. Since the fields of $S_{\rm bTFT}$ are inert under S-duality, it should be $\slz$-invariant as a functional of them, with modular weights arising only in terms of $\tau$.

The holographic partition function provides a sharper test. On $\mathrm{AdS}_5\times S^5$, the pseudo-action vanishes unless a topological five-form term is added~\cite{Kurlyand:2022vzv,Adhikari:2026iib} (see also Ref.~\cite{Chakrabarti:2026uwa}). In our formulation, such a term should arise from first principles. Holography requires $\Mten$ to have a (conformal) boundary, so $\Mel$ acquires a corner $M_9=\partial\Mten$, where the counterterms of holographic renormalization reside and where this term may arise.

Finally, a topological polarization at the other end of an interval would turn $S_{\rm bTFT}$ into a SymTFT for type IIB, which should capture its higher-form and noninvertible symmetries~\cite{Apruzzi:2023uma,Fernandez-Melgarejo:2024ohe}. 

\begin{acknowledgments}
\subsection{Acknowledgments}

We thank Giacomo Giorgi for discussions on the democratic, $\sltwo$-invariant pseudo-action of Ref.~\cite{Fernandez-Melgarejo:2023upq}. This work was supported by the Brain Pool program (RS-2025-25457100) and the Individual Basic Research program (RS-2025-00564305) of the National Research Foundation of Korea (NRF), funded by the Ministry of Science and ICT (MSIT).
\end{acknowledgments}

\end{document}